\documentclass{epl}

\usepackage{latexsym}
\usepackage{epsfig}
\usepackage{amssymb}

\newcommand{\la}{\lambda}
\newcommand{\om}{\omega}

\title{Parametric phase transition in one dimension.}

\author{Jean Farago\inst{1} \And C. Van den Broeck\inst{2}}

\institute{\inst{1} Laboratoire de Physique Statistique, ENS Paris,
75231 Paris cedex 05, France \\
 \inst{2} Limburgs Universitair Centrum, B-3590 Diepenbeek, Belgium}


\begin{document}

\maketitle

\begin{abstract}
We calculate analytically the phase boundary for a nonequilibrium phase
transition in a one-dimensional array of coupled, overdamped parametric
harmonic oscillators in
the limit of strong and weak spatial coupling. Our results show that  the
transition is
reentrant with respect to the spatial coupling in agreement with the
prediction of the mean field theory.
\end{abstract}

Since the landmark paper of Onsager on the solution of the 2-dimensional  Ising
model \cite{onsager}, the search for exactly solvable models displaying
equilibrium phase
transitions has become one of the hot pursuits in statistical physics. Over
the last  two
decades, the interest has shifted from equilibrium to nonequilibrium phase
transitions
\cite{marro}. In many respects, the latter transitions are richer and  more
difficult to
analyse. In particular, the analogue of the equilibrium  distribution is
typically not
known. On the other hand, the arguments ruling out  phase transitions in
one dimension for
a large class of equilibrium models no longer apply, and a number of exact
solutions for
nonequilibrium phase transitions in one dimension have been found
\cite{privman}.  Our
focus here is on a nonequilibrium phase transition in a one-dimensional
array  of coupled
overdamped parametric oscillators. This new transition was  discovered
\cite{vdbkawai}
in the line of work on phase transitions induced by multiplicative  noise
\cite{nit,sancho}, see also \cite{kramer}. The phase transition occurs in
all dimensions.
Surprisingly, it  was found that a mean field analysis predicts quite
accurately the
location of  the phase boundary as obtained from numerical simulations for the
one-dimensional  array. Furthermore, the transition was found to be
reentrant with respect
to the  coupling strength. In this letter, we derive the analytical form of
the phase
boundary in the limit of weak and strong  spatial
coupling. Our exact results allow for a detailed comparison with the the
mean  field
result. In particular the reentrance of the transition is firmly
established. \\

The system considered in \cite{vdbkawai} is a linear chain of overdamped
particles
coupled harmonically and parametrically excited :
\begin{equation}
\dot{x}_i=[-1+A\cos(\omega t+\phi_i)]x_i-K(2x_i-x_{i+1}-x_{i-1}).\label{1}
\end{equation}
The disorder stems from a random choice of the phases $\phi_i$ in
$[0,2\pi]$. This system exhibits  a  nonequilibrium phase transition
from a quiescent phase to an exploding phase at
a   critical  value of the perturbation amplitude $A$. In the exploding
phase the amplitude
of all oscillators diverges with time. This is surprising because if one
suppresses either
the disorder (i.e., one replaces the
$\phi_i$-s by a unique $\phi$) or the coupling (i.e., setting $K=0$), no
destabilization
occurs. Hence, the instability is entirely caused by the conjugate effects of
coupling and disorder.

To analyse the model further, it is convenient to introduce a dimensionless
time unit
$\tau=\omega t$ and scaled variables $\alpha=A/\omega$ and $\kappa=K/\omega$.
Turning first  to the weak coupling limit  $\kappa\ll 1$, we
 switch to new variables $y_i$,
$x_i=y_i\times\exp[-(1/\omega+2\kappa)\tau+\alpha\sin(\tau+\phi_i)]$,
and note that the overall time evolution of these variables:
\begin{equation}
\dot{y}_i=\kappa (e^{\alpha\sin(\tau+\phi_{i+1})-\alpha\sin(\tau+\phi_{i})}
y_{i+1}+e^{\alpha\sin(\tau+\phi_{i-1})-\alpha\sin(\tau+\phi_{i})}y_{i-1})\label{3}
\end{equation}
is slow due to the  assumption that the prefactor $\kappa$ in the
r.h.s. of (\ref{3}) is
small. On this slow  time scale the periodic
parametric perturbations  are fast and one can   replace in (\ref{3})  the
following quantities
by their average over one period of the oscillations:
\begin{eqnarray}
e^{\alpha\sin(\tau+\phi_{i+1})-\alpha\sin(\tau+\phi_{i})}&\leadsto&
\frac{1}{2\pi}\int_0^{2\pi}d\tau\
e^{\alpha\sin(\tau+\phi_{i+1})-\alpha\sin(\tau+\phi_{i})} \nonumber \\
&=& \frac{1}{2\pi}\int_0^{2\pi}d\tau \
e^{2\alpha\sin(\frac{\phi_{i+1}-\phi_i}{2})\cos
\tau}
 \equiv \lambda_i.\label{3.1}
\end{eqnarray}
This procedure is tantamount to the Bogoliubov-Mitropolski technique,
familiar from the literature on oscillators \cite{nayfeh}. With  the
 replacement given in (\ref{3.1}),  (\ref{3}) reduces to :
\begin{equation}
\dot{y}_i=\kappa (\lambda_i y_{i+1} +\lambda_{i-1}y_{i-1}).
\end{equation}
Clearly, for almost all initial conditions,  the asymptotic time dependence
of $y_i$ will
be dominated by the largest eigenvalue $\lambda_+$ of the symmetric
tri-diagonal random matrix
$
\Lambda_{i,i+1}=\Lambda_{i+1,i}=\lambda_i,\ \ \forall i
$, the remaining elements $\Lambda_{i,j}$ being zero. The phase boundary
for the transition
to explosion is thus given by
$(1/\omega+2\kappa)/{\kappa}=\lambda_+$.

\begin{figure}[h]
\onefigure[scale=0.4]{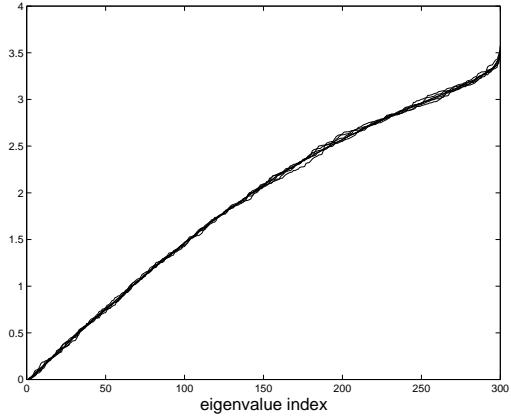}
\caption{Positive part of five realizations of $600\times600$ symmetric
random matrices ; these
matrices are of the $\Lambda$-type, that is tri-diagonal symmetric with
zero elements on
the diagonal, and random independent  samplings
in the interval
$[1,2]$ on the off-diagonal line. The upper bound of the main spectrum {\it
appears} to be
strictly smaller than
$4$, in disagreement with the more careful analysis given in the
appendix.}\label{fig1}
\end{figure}

In the limit of an infinite system, the value of $\lambda_+$ is expected to be
self-averaging. To determine this value, we note
that
  the $\lambda_i$'s are statistically
independent, hence our problem belongs to the traditional theory of random
tri-diagonal symmetric matrices with independent random elements. Since in
our case the
diagonal elements of the matrix
$\Lambda$ are zero - unlike the famous cases of matrices coming from
disordered  models of
linear chains with randomly chosen  spring constants and/or masses -
our matrix belongs to the ``type I'' disorder in Dyson's terminology
\cite{dyson,mattis}. Furthermore the support of
$\lambda_i$ is bounded, $\lambda_i\in[\lambda_{min},\lambda_{max}]$, and
 a theorem due to Hori and Matsuda \cite{horimatsuda, mattis} states that
$\lambda_+$ is bounded from above by $2\lambda_{max}$. The latter bound is
realized for
the  choice $\lambda_i=\lambda_{max}, \forall i$. We are however concerned
with the
typical value of $\lambda_+$, as it is observed with probability one in the
thermodynamic
limit. A numerical evaluation of the spectrum of a large $\Lambda$
matrix in fact points to a value of $\lambda_+$ significantly lower than
the "naive"
estimation given by the upperbound (cf. fig. \ref{fig1}). A detailed
analytic analysis (cf.
appendix for the major steps of the proof) however
reveals that  $2\lambda_{max}$ is in fact the typical value of  $\lambda_+$
and the observed
deviations in the numerics are entirely the result of strong finite-size
effects.

We conclude that in the limit of small $\kappa$  the phase boundary is
given by:
\begin{equation}
\frac{1/\omega
+2\kappa}{\kappa}=2\lambda_{max}=\frac{2}{\pi}\int_{0}^\pi d\tau\
e^{2\alpha\cos \tau} =2I_0(2\alpha).\label{10}
\end{equation}
This result has to be compared with the mean-field prediction \cite{vdbkawai}:
\begin{equation}
\int_0^\infty d\tau
e^{-(1/\omega+\kappa)\tau}I_0(2\alpha\sin(\tau/2))=\frac{1}{\kappa},\label{11}\end{equation}
which, in the region of $\kappa$ small and $\alpha$ large, reduces to:
\begin{equation}
\frac{I_0(2\alpha)}{\sqrt{\alpha}}=
\frac{\sinh[\pi(1+1/\omega)]}{\sqrt{\pi}\kappa}.
\end{equation}
The two asymptotic formulae are not identical, but the leading term of the
asymptotic branch is
the same, namely $\alpha\sim -\frac{1}{2}\log\kappa$. This partly explains
the observed similarity of the two instability curves in this region.
 We close the discussion of the small $\kappa$ limit
 by noting that the above method can unfortunately not be generalized to
higher dimensions : the arguments
are very specific to dimension 1, while the variables $\lambda_i$ are no
longer independent in higher dimensions.

\bigskip

Next we turn to the description of the large $\kappa$ regime.
In this case it is convenient to transform away  the $-x_i$ term by setting
$x_i=y_i\exp(-\tau/\omega)$, and to introduce the discrete Fourier transforms:
\begin{eqnarray}
y_q&\equiv&\frac{1}{\sqrt{N}}\sum_{j=0}^{N-1}y_je^{2i\pi jq/N},\ \ \
q=0,\ldots,N-1.
\end{eqnarray}
The evolution equation for these Fourier modes reads :
\begin{eqnarray}
\dot{y}_q &=& \alpha\sum_{q'=0}^{N-1}
y_{q'}C_{q-q'}(\tau)-\left(4\kappa\sin^2\frac{\pi q}{N}\right)y_q, \label{14}\\
\mbox{with}\ \ C_q(\tau)&=&\frac{1}{N}\sum_{j=0}^{N-1}e^{2i\pi
jq/N}\cos(\tau+\phi_j).\nonumber
\end{eqnarray}
In the limit of
$\kappa$ large, the system becomes very "stiff" and any initial
short wavelength perturbation is expected to be damped out very rapidly,
cf. second term in
the r.h.s. of (\ref{14}).
 We therefore assume that, as can be verified a posteriori,  the long-time
behaviour of the
$q\neq0$ modes is completely determined by their coupling to the slow
zero-th mode $q=0$, i.e.
by the "reinjection" of energy through the $q'=0$ contribution in the first
term of the r.h.s.
of (\ref{14}).
Dropping furthermore the irrelevant contributions due to
initial conditions, one concludes that for $q\neq 0$:
\begin{equation}
y_q=\alpha\int_0^t du e^{-[4\kappa\sin^2\pi q/N](t-u)}y_0(u)C_q(u).
\end{equation}
Inserting this result in  the equation for $q= 0$  leads to:
\begin{equation}
\dot{y}_0=\alpha^2 \sum_{q=0}^{N-1}\int_0^t du e^{-[4\kappa\sin^2\pi
q/N](t-u)}y_0(u)C_q(u)C_{-q}(t).
\end{equation}
One expects that the quantity $\sum_{q=0}^{N-1}Ne^{-[4\kappa\sin^2\pi
q/N](t-u)}C_q(u)C_{-q}(t)$ is self-averaging in thermodynamic limit
\cite{comment}, hence one can
perform an average over the phase disorder directly in the dynamical
equation:
\begin{eqnarray}
\dot{y}_0&=&\frac{\alpha^2}{2} \int_0^t du
\left(\frac{1}{N}\sum_{q=1}^Ne^{-[4\kappa\sin^2\pi
q/N](t-u)}\right)\cos(t-u)y_0(u)\\
&=&\frac{\alpha^2}{2} \int_0^t du
\left(\int_0^1dq e^{-[4\kappa\sin^2\pi
q](t-u)}\right)\cos(t-u)y_0(u).
\end{eqnarray}
where the thermodynamic limit has been taken in the second equality.
The resulting
equation
can be readily solved for the Laplace transform $\widehat{y}_0(z)$
 of $y_0$ (assuming $y_0(t\!=\!0)=1$) :
\begin{equation}
\widehat{y}_0(z)=\frac{1}{z-\frac{\alpha^2}{2}\widehat{g}(z)},
\end{equation}
where $\widehat{g}(z)$, the  Laplace transform of $g(t)=\cos t\times\int_0^1dq\
e^{-[4\kappa\sin^2\pi q]t}$, is found to be
\begin{equation}
\widehat{g}(z)=\left(\frac{\sqrt{(z^2+1)((z+4\kappa)^2+1)}+z(z+4\kappa)-1}
{2(z^2+1)((z+4\kappa)^2+1)}\right)^{\frac{1}{2}}.
\end{equation}
The long-time asymptotic behaviour of $y_0(t)$ turns out to be governed
 by the single real positive pole $\gamma$ of $\widehat{y}_0(z)$. The
 phase transition to explosion takes place when $\gamma=1/\omega$, or more
explicitly when
\begin{equation}
\frac{1}{\omega}=\frac{\alpha^2}{2}\widehat{g}(1/\omega)\simeq
\frac{\alpha^2}{4\sqrt{2\kappa}}
\left(\frac{\sqrt{\omega^{-2}+1}+\omega^{-1}}{\omega^{-2}+1}
\right)^\frac{1}{2},
\end{equation}
where the large $\kappa$ limit is invoked in the second equality.
The dependence $\alpha\propto \kappa^{1/4}$ explains the flat shape of the
reentrance curve.  In comparison the mean-field
model predicts a behaviour $\alpha\sim\sqrt{2\kappa/\omega}$, consistent
with
the fact that the
reentrance is expected to occur at a smaller value of $\kappa$ (for a given
value of $\alpha$) in
the mean-field system, since the connectivity of the system enhances its
effective stiffness.
\\

\begin{figure}
\onefigure[scale=0.4]{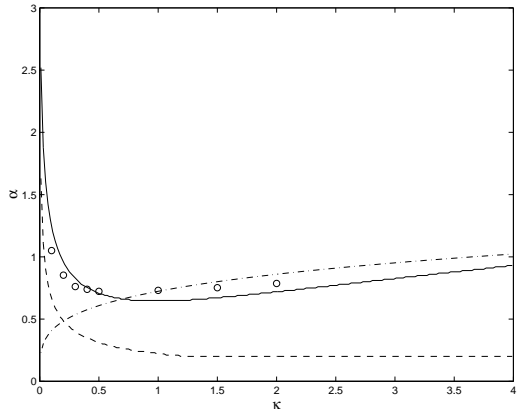}
\caption{Comparison between stability curves for the mean-field model
(solid line),
1-d prediction for small  (dashed line) and high
(dash-dotted line) $\kappa$ values. The numerical results of
\cite{vdbkawai} are also recalled (circles) (for all these curves, $\om=10$).}
\label{fig2}
\end{figure}

We finally turn to a comparison of the above derived asymptotic results
with the mean field and
numerical results, cf.  fig. \ref{fig2}.
 The differences between the asymptotic 1D results and
 the mean-field value are rather small. More revealing is that
the 1D system is less stable than the mean-field model for small and (very)
large $\kappa$
values (cf. the respective asymptotics $ \kappa^{1/4} \ll \kappa^{1/2}$),
while
the opposite seems to be the case  for intermediate values :
 if the asymptotic evaluation for $\kappa$ large is
reliable for $\kappa\sim 4$, we can see from fig. \ref{fig2} that there exists
  a range of coupling values for which  the 1D system is
more stable than the mean-field model. This inversion seems to be confirmed by
the  numerical
simulations of the 1D model  in \cite{vdbkawai} and  highlights
the subtleties of the destabilisation process taking place in the system.

 We conclude that the analytic theory
presented here describes
the salient features of the phase boundary for the nonequilibrium phase
transition occurring in a
one-dimensional array of coupled, overdamped parametric  harmonic
oscillators with quenched
random phases. The fact that the deviations from the mean field results are
small raises the
question of the critical dimension.
Interestingly, the above presented procedure to calculate the phase
boundary for large
$\kappa$ can also be applied to  higher dimensions. One finds a mean field
behaviour for a
spatial dimension larger or equal to $3$, with the reentrance curve given by
$\alpha\propto\sqrt{\kappa}$. The dimension $2$ is ``critical'', characterized
by logarithmic corrections $\alpha\propto\sqrt{\kappa/\log\kappa}$. This
strongly suggests
that the critical dimension of the phase transition is in fact equal to two.

\section{Appendix}

In this appendix, we review some spectral properties of the random
matrix $\Lambda$. First, one notices that its spectrum is always
symmetric with respect to zero, for if $\{y_n\}$ is an eigenvector associated
 with the eigenvalue $\la$, then $\{(-1)^ny_n\}$ is another
eigenvector with the eigenvalue $-\la$. We can thus restrict our attention
to the negative eigenvalues $-\la,\la>0$. The eigenvector
equation reads $-\la y_n=\la_ny_{n+1}+\la_{n-1}y_{n-1}$ and can be
recast into the form $z_{n+1}=-\mu_n/(\la+z_n)$, with
 $z_n=\la_{n-1}y_{n-1}/y_n,\mu_n=\la_n^2$. One can
now reproduce step by step  the demonstration performed by Schmidt
\cite{schmidt},
whose conclusions are that the proportion of negative  eigenvalues comprised in
$[-\la,0]$ is equal in the thermodynamic limit to the proportion of
negative $z_n$ in the sequence $z_0,z_1,\ldots$. Turning first to the
case where the $\mu_n$ are constant (=$\mu$) two different
situations arise. For $\la^2<4\mu$, the mapping from $z_n$ to $z_{n+1}$ has
 no (real) fixed points, and as a result the positive $z_n$ are substantially
visited (i.e. the proportion of positive $z_n$ is non zero when
$N\rightarrow\infty$).
 For  $\la^2>4\mu$
the appearance of  a
 globally attractive negative fixed point $z_+=(-\la+\sqrt{\la^2-4\mu})/2$
 implies that the sequence will ultimately end up in the
$\mathbb{R}^-$ region.
 Hence, $-\la=-2\sqrt{\mu}$ is
the lower bound of the spectrum, since beyond this value, the
proportion of negative $z_n$-values is $1$.

We next  turn to the more complicated case of $\la_n$ being independent random
variables sampled from a distribution with support
$[\la_{min},\la_{max}]$ (with $\la_{min}\geqslant 0$). Since the
$\mu\in[\mu_{min},\mu_{max}]$ are also independent random
variables, the sequence $z_n$ becomes a Markov process.
In continuation of the arguments given above,
the following conclusions can be drawn.
For  $\la>2\la_{max}$ it follows that  $\la^2>4\mu_n$ is true for all
allowed values
of $\mu_n$, and the $z_n$ are attracted to a
region $[z_+(\mu_{max}),z_+(\mu_{min})]$ which lies entirely on the negative
real axis. One concludes that the spectrum, assumed to be
 self-averaging in the limit $N \rightarrow \infty$,
cannot extend beyond the value $2\la_{max}$ (a result also
demonstrated in \cite{horimatsuda}). Considering the situation where
$\la$ is slightly less than $2\la_{max}$, the $z_n$ will still more often
than not be attracted
to the region $[-\la/2, z_+(\mu_{min})]$. Exceptionally however, when $\mu_n$
is sampled outside $[\mu_{min},\la^2/4]$,  the  $z_n$  will make an excursion
toward the region $]-\infty,-\la]$.
If a long
enough such series of successive $\mu_n$ is sampled,  $z_n$
will eventually cross the $-\la$ value and next become positive. Even
though  unlikely,
the probability for these events is not zero in
thermodynamic limit, for their occurrence only requires a finite number
of steps in the sequence of $z_n$. As a result, one concludes that
the effective boundary for the spectrum is indeed $\pm
2\la_{max}$. Moreover, one can understand qualitatively why the spectra
of figure \ref{fig1} do not seem to saturate this bound as follows.
The number of successive steps with
$\mu_n\in I=[\la^2/4,\mu_{max}]$,  required to reach $\mathbb{R}^+$ can
be considered more or less as constant (with respect to $\la$), say
$n_0$. Then the probability for observing a positive value of $z_n$ is
of order $p^{n_0}$, where $p$ is the probability to sample $\mu_n$ in
$I$. In the example of figure \ref{fig1}, $p$ is simply given by
$(\lambda_{max}-\la/2)/(\lambda_{max}-\lambda_{min})=2-\la/2$.
Moreover, a rough
evaluation of $n_0$ gives \textit{at least} $4$ steps. Then the
proportion of eigenmodes greater than $3.5$ can be (very generously !)
estimated to be
$(2-3.5/2)^4=0.3\%\ldots$, explaining  the
apparent absence of these eigenvalues in figure \ref{fig1}.


\begin{thebibliography}{99}
\bibitem{comment} this includes the assumption that no macroscopic
correlations appear
between this quantity and $y_0$.
\end{thebibliography}
\end{document}